\documentclass{iau} 
\usepackage{graphicx}

\title[Modelling the outskirts of galaxies in a cosmological context] 
{Modelling the outskirts of galaxies\\in a cosmological context}

\author[Andrew P. Cooper]   
{Andrew P. Cooper$^1$}

\affiliation{$^1$Institute for Computational Cosmology, Durham University \\ Science Site, South Road, Durham, DH1 3LE, UK \\ email: {\tt a.p.cooper@durham.ac.uk} }

\pubyear{2017}
\volume{321}  
\jname{Formation and evolution of galaxy outskirts}
\editors{A. Gil de Paz, J. C. Lee \& J. H. Knapen, eds.}
\begin{document}

\maketitle

\begin{abstract}


  Current data broadly support trends of galaxy surface brightness profile
  amplitude and shape with total stellar mass predicted by state-of-the-art
  $\Lambda$CDM cosmological simulations, although recent results show signs of
  interesting discrepancies, particularly for galaxies less massive than the
  Milky Way. Here I discuss how perhaps the largest contribution to such
  discrepancies can be inferred almost directly from how well a given model
  agrees with the observed present-day galaxy stellar mass function.


  \keywords{galaxies: evolution; galaxies: halos; galaxies: structure; methods: n-body simulations}
\end{abstract}


Many remarkable observations of galaxy structure to extremely low surface
brightness levels are discussed in these proceedings.  These observations and
quantities derived from them, such as `stellar halo' mass, can be compared to
predictions from modern cosmological simulations of galaxy formation. Such
comparisons test a generic prediction that the amplitude and shape of a
galaxy's surface brightness profile correlate closely with its stellar mass and
even more closely with the mass of its host dark matter (DM) halo.  

That prediction is founded on some of the most fundamental aspects of the
theory of galaxy formation in a cold dark matter (CDM) universe.  Galaxies form
through the dissipative collapse of the gas trapped by the potential wells of
DM halos (\cite[White \& Rees 1978]{White:1978aa}; \cite[White \& Frenk
1991]{White:1991aa}).  Hierarchical clustering of halos leads to the accretion
of additional stellar mass through mergers (see \cite[Guo \& White
2008]{Guo:2008aa} for a comparison of the relative contributions of merging and
\textit{in situ} growth to galaxies and DM halos).  If the galaxy formation
process was self-similar, the galactic stellar mass function would simply be a
scaled-down version of the $\Lambda$CDM halo mass function.  Observed
luminosity functions, however, have a significantly different shape.  The
`efficiency' of galaxy formation (i.e.  the fraction of baryonic mass locked in
long-lived stars per unit total mass) must therefore vary with halo mass.  Much
of the complexity of galaxy formation lies in identifying the processes (and
interactions) that modulate this variation (e.g.  \cite[{Kauffmann}, {White},
\& {Guiderdoni} 1999]{Kauffmann:1993aa}; \cite[{Cole} \etal\
2000]{Cole:2000aa}; \cite[{Bower} \etal\ 2006]{Bower06}).

\begin{figure}[t]
\begin{center}
\includegraphics[width=62mm, clip=True, trim=1mm 2mm 1mm 2mm ]{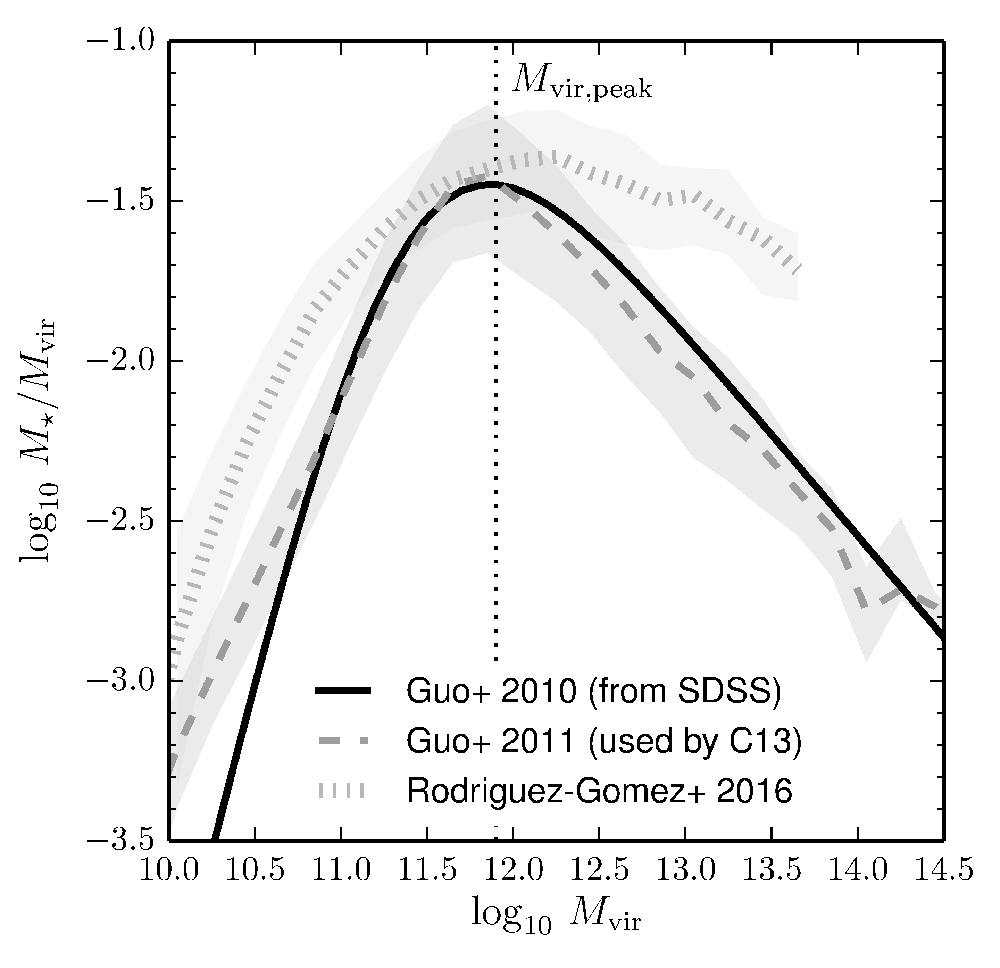} 
\includegraphics[width=60mm, clip=True, trim=1mm 2mm 1mm 2mm ]{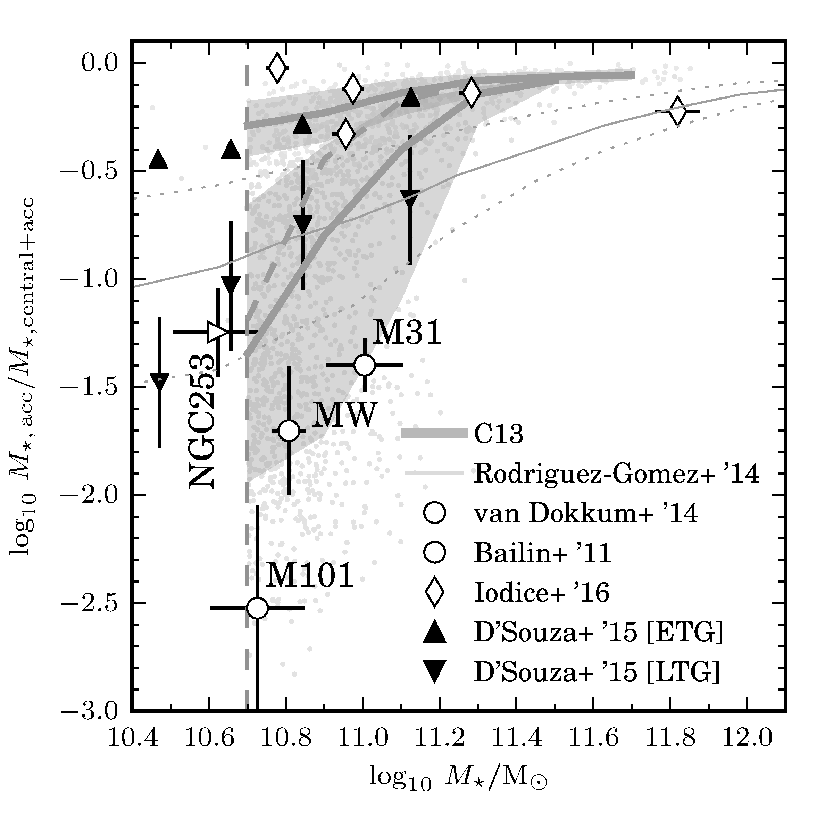} 

\caption{\textit{Left:} present-day ratio of galaxy stellar mass to total mass
(see text) as a function of stellar mass. The \cite{Guo:2010aa} SDSS abundance
matching result is compared to predictions of the \cite{Guo:2011ab} model used
by C13 and the Ilustris simulation used by \cite[{Rodriguez-Gomez} \etal\
(2016)]{Rodriguez-Gomez:2016aa} (both use the SDSS mass function as a
constraint). Note the log scale. \textit{Right}: Mass fraction in the accreted
stellar halos of galaxies vs. their stellar mass. Shaded bands bracket the
$16$-$84^{th}$ percentile range from C13 for early (upper) and late types
(lower; split at $B/T = 0.7$); solid lines are medians. The thick dashed line
is the median of \textit{all} galaxies in C13 and the vertical dashed line the
mass limit of their sample. Dotted lines bracket the same range of the
predictions from Illustris.  Symbols show observational results (see text).
Open symbols are individual galaxies, noted in legend.  Filled triangles are
SDSS stacks of \cite[d'Souza \etal\ (2014)]{DSouza:2014aa} split into early and
late types by light profile concentration $C=R_{90}/R_{50}$.}

\label{fig:1}
\end{center}
\end{figure}

Under a simple, well motived set of assumptions, the so-called abundance
matching technique can be used to estimate the trend of galaxy formation
efficiency with halo mass directly from the observed galaxy stellar mass
function \cite[(e.g. Guo \etal\ 2010)]{Guo2010:aa}.  One such estimate is
illustrated in the left-hand panel of Fig.~\ref{fig:1}, together with two
theoretical predictions based on forward modelling. Galaxy formation efficiency
peaks around the likely halo mass of the Milky Way (MW).  It declines at low
halo mass due to the effectiveness of supernova feedback in ejecting gas from
shallow potentials, and at high masses (according to current understanding) due
to the onset of AGN feedback.  Recognizing that the mass ratio distributions of
accreted DM halos are approximately self-similar for all host halo masses and
assuming, as a simplification, that this relationship does not evolve over
time, some straightforward inferences about how the mass (fraction) of stars
accreted by galaxies varies with their halo mass then follow.

Most satellites accreted by a host halo with mass $M_{\mathrm{host}} \le
M_{\mathrm{vir, peak}}$ (see Fig.~\ref{fig:1}) will have mass-to-light ratios
one or more orders of magnitude higher than the host itself. The stellar halos
of these galaxies will be dominated by the one or two most massive (hence most
recently disrupted) progenitors.  The accretion times and orbits of individual
progenitors are highly stochastic and the intrinsic scatter of \textit{in situ}
star formation efficiency may also increase at lower mass. Hence we expect a
large dispersion in the mass fraction of accreted stars for $M_{\mathrm{host}}
\le M_{\mathrm{vir, peak}}$. The are other implications; a significant
difference in metallicity between deeply embedded \textit{in situ} stars and
diffuse accreted stars should give rise to a steep (apparent) gradient in
metallicity for the host halo. Conversely, since \textit{in situ} star
formation is suppressed at $M_{\mathrm{host}} \ge M_{\mathrm{vir, peak}}$, the
most massive satellites accreted by those hosts will have roughly similar
stellar masses over a wide range of halo mass. \cite[Cooper \etal\ (2013
$\mathrm{[C13]}$; 2015 $\mathrm{[C15]}$)]{Cooper:2013aa, Cooper:2015ab} found
that typical cluster BCGs are built by $\sim10$ equally significant
contributors.  Such systems are dominated throughout by their accreted
spheroids, with consequently high Sersic index, small effective radius, narrow
halo-to-halo scatter and homogeneous stellar populations \cite[(e.g. {De Lucia}
\& {Blaizot} 2007)]{DeLucia:2007aa}.

\cite[Purcell, Bullock \& Zentner (2007)]{Purcell:2007aa} were the first to
make explicit predictions for accreted stellar mass fractions over a wide range
of halo mass based on these ideas. More recently the problem has been tackled
in greater detail by forward modelling with cosmological simulations: here we
consider C13 and \cite[{Rodriguez-Gomez}, \etal\ (2016
$\mathrm{[Illustris]}$)]{Rodriguez-Gomez:2016aa}. The right-hand panel of
Fig.~\ref{fig:1} compares predicted stellar halo mass fractions from these two
simulations against various observational results, as a function of total
\textit{stellar} mass (as a more readily observable albeit noisy proxy for
$M_{\mathrm{host}}$). I take all these results at face value as being directly
comparable to to the (unambiguously defined) total accreted stellar mass in the
simulations.  However, as discussed elsewhere in this volume, the observations
use a variety of techniques which are not easy to compare even with one
another. Moreover there is a fraught problem of definition (discussed in C13
and C15): generally the `accreted mass' is inferred by extrapolating fits to
data from the outskirts of galaxies inwards.  The implied distinctions between
accreted and \textit{in situ} stars in the `inner' galaxy differ greatly from
one study to the next. An explicit, objective definition of `the stellar halo'
for a particular set of observations is important to ensure a robust comparison
with models. A study carefully comparing observational inferences of stellar
halo mass in galaxies of different types with different techniques would be
very useful.  

In Fig.~\ref{fig:1}, the Illustris relation is notably shallower than that of
C13, with similar accreted fraction around the MW mass but much higher
fractions (and smaller scatter) at lower masses, and vice versa.  Comparing the
$M_{\star}/M_{\mathrm{vir}}$ relations in the left-hand panel, it is clear that
Illustris predicts similar star formation efficiency around $M_{\mathrm{vir,
peak}}$ but associates substantially more stellar mass with both low- and
high-mass halos (presumably through enhanced \textit{in situ} star formation).
Following the logic above, this readily explains the different predictions of
the accreted stellar mass fraction.  The large scatter at low $M_{\star}$ is
very clear in the observations (more evidence of this is provided in other
contributions).  Also clear is the separation between early and late types, at
least in the SDSS stacks of \cite{DSouza:2014aa}.  Curiously, the MW and M31
both appear well below the stacked averages for late types of similar mass,
albeit with the $1\sigma$ contour of C13. At the high mass end, problems of
definition become increasingly important, because it is no longer possible to
distinguish two major components in the light profile. Fitting to smaller
perturbations in the outskirts (see e.g. Fig. 10 of C15) can lead to `halo'
mass fractions much lower than the true fraction of accreted stellar mass.  

In summary, the galaxy stellar mass function is a simple but extremely
important point of comparison when interpreting discrepancies between
theoretical models of stellar halos, as it is when using models to interpret
observational data. Taking the issues above into account, future systematic
surveys of diffuse light may be able to constrain the more subtle trends in the
structure of galaxy outskirts that emerge, according to cosmological
simulations, from the complex co-evolution of galaxies and their DM halos.


\begin{thebibliography}{}

  \bibitem[{Bower} \etal\ (2006)]
  {Bower06}
  {Bower}, R. G., \etal\ 2006, {\em MNRAS, }370, 645-655
\bibitem[{Cole} \etal\ (2000)]{Cole:2000aa}
  {Cole}, S., {Lacey}, C. G., {Baugh}, C. M., \& {Frenk}, C. S. 2000, {\em MNRAS, }319, 168-204
\bibitem[{Cooper} \etal\ (2013)]
  {Cooper:2013aa}
  {Cooper}, A. P., \etal\ 2013, {\em MNRAS, }434, 3348-3367; C13
\bibitem[{Cooper} \etal\ (2015)]
  {Cooper:2015ab}
  {Cooper}, A. P., \etal\ 2015, {\em MNRAS, }451, 2703-2722
\bibitem[{De Lucia} \& {Blaizot} (2007)]
  {DeLucia:2007aa}
  {De Lucia}, G.  \& {Blaizot}, J. 2007, {\em MNRAS, }375, 2-14
\bibitem[{D'Souza} \etal\ (2014)]
  {DSouza:2014aa}
  {D'Souza}, R., {Kauffman}, G., {Wang}, J., \& {Vegetti}, S. 2014, {\em MNRAS, }443, 1433-1450
\bibitem[{Guo} \& {White} (2008)]
  {Guo:2008aa}
  {Guo}, Q.  \& {White}, S. D. M. 2008, {\em MNRAS, }384, 2-10
\bibitem[{Guo} \etal\ (2010)]
  {Guo:2010aa}
  {Guo}, Q., {White}, S., {Li}, C., \& {Boylan-Kolchin}, M. 2010, {\em MNRAS, }404, 1111-1120
\bibitem[{Guo} \etal\ (2011)]
  {Guo:2011ab}
  {Guo}, Q., \etal\ 2011, {\em MNRAS, }413, 101-131
\bibitem[{Iodice} \etal\ (2016)]
  {Iodice:2016aa}
  {Iodice}, E. \etal\ 2016, {\em ApJ, }820, 42
\bibitem[{Kauffmann}, {White}, \& {Guiderdoni} (1993)]
  {Kauffmann:1993aa}
  {Kauffmann}, G., {White}, S. D. M., \& {Guiderdoni}, B. 1993, {\em MNRAS, }264, 201
\bibitem[{Purcell}, {Bullock}, & {Zentner} (2007)]
  {Purcell:2007aa}
  {Purcell}, C. W., {Bullock}, J. S., \& {Zentner}, A. R. 2007, {\em ApJ, } 666, 20-33
\bibitem[{Rodriguez-Gomez} \etal\ (2016)]
  {Rodriguez-Gomez:2016aa}
  {Rodriguez-Gomez}, V.,  \etal\ 2016, {\em MNRAS, }458, 2371-2390
\bibitem[{Schaye} \etal\ (2015)]
  {Schaye:2015aa}
  {Schaye}, J., \etal\ 2015, {\em MNRAS, }446, 521-554
\bibitem[{van Dokkum}, {Abraham}, \& {Merritt} (2014)]
  {van-Dokkum:2014aa}
  {van Dokkum}, P. G., {Abraham}, R., \& {Merritt}, A. 2014, {\em ApJL, }782, L24
\bibitem[{White} \& {Rees} (1978)]
  {White:1978aa}
  {White}, S. D. M.  \& {Rees}, M. J. 1978, {\em MNRAS, }183, 341-358
\bibitem[{White} \& {Frenk} (1991)]
  {White:1991aa}
  {White}, S. D. M.  \& {Frenk}, C. S. 1991, {\em ApJ, }379, 52-79

\end{thebibliography}
\end{document}